\documentclass[prb]{revtex4}

\usepackage{graphicx}
\usepackage{dcolumn}
\usepackage{bm}

\def\beq{\begin{equation}}
\def\eeq{\end{equation}}
\def\bea{\begin{eqnarray}}
\def\eea{\end{eqnarray}}

\begin{document}

\title{Ground states of quantum kagom\'e antiferromagnets \\
in a magnetic field}

\author{D.C.\ Cabra}
\affiliation{Laboratoire de Physique Th\'eorique, Universit\'e Louis Pasteur,
3 rue de l'Universit\'e, F-67084 Strasbourg Cedex}

\author{M.D.\ Grynberg}
\affiliation{Departamento de F\'{\i}sica, Universidad Nacional de La Plata \\
C.C.\ 67, (1900) La Plata, Argentina.}

\author{P.\ Pujol}
\affiliation{Laboratoire de Physique, ENS Lyon, 46 All\'ee
d'Italie, 69364 Lyon C\'edex 07, France.}

\begin{abstract}
We study the ground state properties of a quantum antiferromagnet
in the kagom\'e lattice in the presence of a magnetic field,
paying particular attention to the stability of the plateau at
magnetization $1/3$ of saturation. While the plateau is reinforced
by certain deformations of the lattice, like the
introduction of structural defect lines and against an Ising
anisotropy, ground state correlations are seen to be quite different
and the undistorted $SU(2)$ case appears to be rather special.
\end{abstract}

\maketitle


The interest in the study of quantum two-dimensional frustrated
antiferromagnets has been boosted by recent high magnetic field
experiments where different exotic situations have been observed.
Generically, plateaux and jumps in the magnetization curve show up
in materials realizing different frustrated structures \cite{tri}
\cite{j1j2} \cite{SS} \cite{kago}. In the case of the kagom\'e
lattice, the nature of the ground  state at zero field shows
interesting features of a disordered spin liquid with a spin gap
and a macroscopic number of low lying (spinless) excitations
\cite{Lhuillier,Mila}. There is general consensus in that this
lattice shows a clear plateau at magnetization $1/3$ of saturation
\cite{nos} and that in this case the spin gap is larger than in
the zero field case, making it more stable and potentially
observable.

Prompted by these findings, we studied the antiferromagnetic (AF)
Heisenberg model defined on the kagom\'e lattice focusing at the
spinful state with $M=1/3$. We have allowed for different
couplings along the different directions in the kagom\'e lattice
and found that the plateau at $1/3$ survives this distortion and
it is even enhanced. On the other hand, the ordering of the GS
changes substantially at the isotropic point, whose nature is
still under debate \cite{horda}.

Interestingly, our results also reveal the emergence of a whole
regime ($J'/J < 0.62\,$), in which the {\it spontaneous}
susceptibility diverges at $M = 1/3\,$, i.e.\ all lower fields
vanish $\forall \,\, M \le 1/3\,$. Recent magnetic field
measurements in the novel cuprate Volborthite,
Cu$_3$V$_2$O$_7$(OH)$_2$-2H$_2$O \cite{Vol}, which is apparently
described by a distorted kagom\'e lattice like the one studied in
the present paper, show no plateau at zero magnetization down to
1.8K. This could be taken as an indication that this material lies
within the region of couplings slightly above $J'/J \sim 0.62$,
where fairly small spin excitations are observed. If this turns
out to be the case, this would imply that the predicted $1/3$
plateau could be accessible to the magnetic fields currently
available, since this plateau opens up for fairly small field
values within this range of couplings.

Turning to the GS orderings, it is important to emphasize that all
studied samples at $M=1/3\,$ yield a rather small non-magnetic gap
($\Delta S^z = 0\,$) which is at least one order of magnitude
smaller than the involved spin exchange interactions. We also
observed that for $J'/J = 3/4\,$ the spin pair correlations
$\langle \,\sigma^z_0 \sigma^z_{\bf r} \, \rangle\,$ display long
distance ferromagnetic (F) order along the $J'\!$-direction
whereas the $J$-directions become clearly antiferromagnetic (AF),
as is shown in Fig.\ 3 (We call this hereafter AF-F ordering). For
$J'/J > 1\,$ these AF-F orderings directions result exchanged.

A completely different scenario arises when $J=J'$, where though
non-magnetic excitations still remain very small, the radial
average over pair correlation functions resembles the
characteristics of a  $\sqrt 3 \times \sqrt 3$ spin liquid rather
than those of an ordered state. However, further work is needed to
understand the GS structure in the isotropic $SU(2)$ case
\cite{horda}.

We have also studied the overlaps of the GS wave function with the
real space $\sigma^z\!$-configurations. Specifically, Fig.\ 3
displays the probabilities $\vert\,\langle s \vert \psi\,\rangle\,
\vert^{\,2}\,$ of these $\vert \, s \,\rangle$-states after
sorting their weights on the $\vert\, \psi\,\rangle\,$ GS  of a
given cluster. The key issue here is that a clean probability
scale separation -more than one order of magnitude-, shows up
between the AF-F configuration and the rest of the GS components.
However, it should be stressed that the point $J=J'\,$ is special
in that it exhibits an entirely different scenario. In that case
the hierarchy of probability scales is smeared out completely and
a much larger number of states seems to intervene on the pair
correlations. The point here is that the maximum GS overlap can
therefore be regarded as an alternative measure of order, so long
as the hierarchical overlap distribution still holds.


We acknowledge useful discussions with A.\ Honecker and P.C.W.\
Holdsworth.





\begin{figure}
\centerline{\includegraphics[width=90mm]{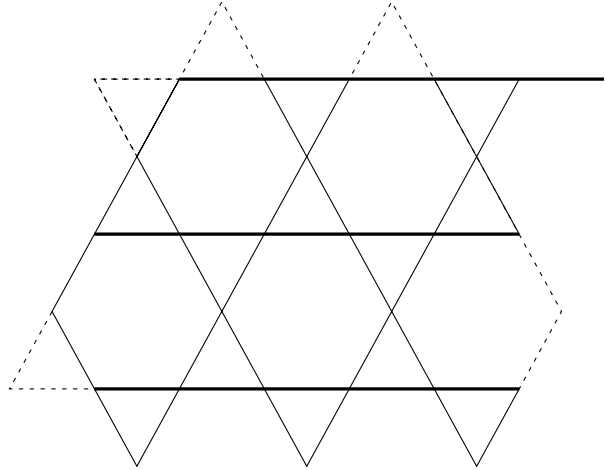}}
\caption{Schematic depiction of the type of cluster considered in
exact diagonalizations. Bold horizontal lines denote
$J'$-exchanges whereas thin solid lines stand for $J$-exchanges.
Dotted lines stand for periodic boundary conditions which in turn
can tile the lattice in different periodic ways. }
\end{figure}

\begin{figure}
\vskip -2cm \centerline{\includegraphics[width=100mm]{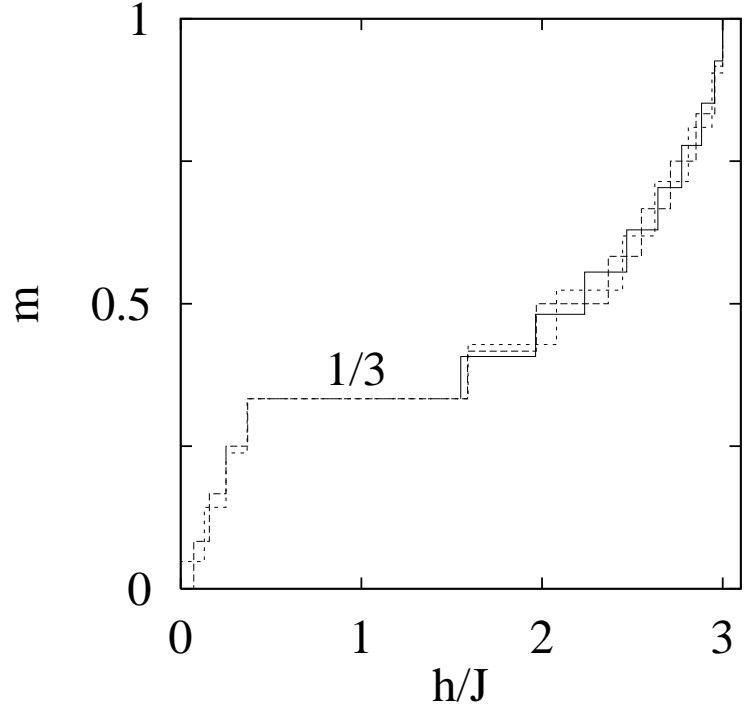}}
\caption{Magnetization curves of different cluster sizes $N$
obtained for $N = 21$ (dotted lines), 24 (dashed lines) and 27
spins (solid lines), using $J'/J = 3/4$.}
\end{figure}

\begin{figure}
\vskip -8cm \centerline{\includegraphics[width=130mm]{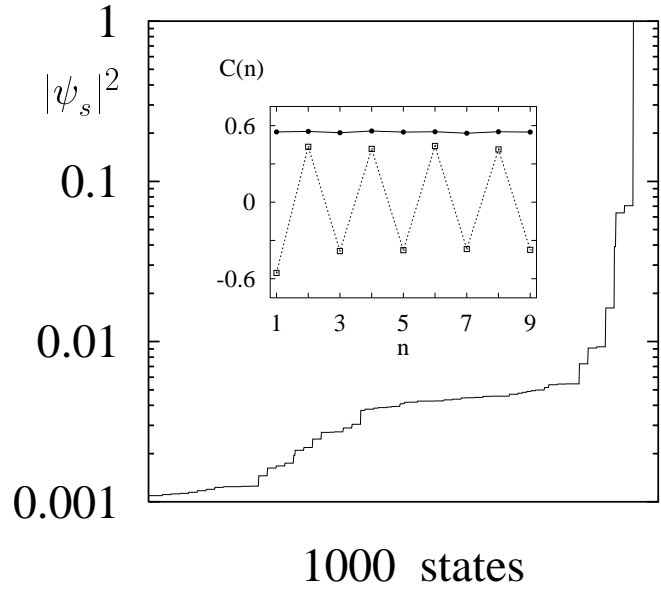}}
\caption{Ground state for $J'/J = 0.9$ at $M = 1/3$ for $N=27$.
Real space states in the $\sigma^z$-representation are sorted by
increasing overlaps and normalized to their maximum value (only
overlaps above $10^{-3}$ are shown). The latter corresponds to and
AF-F state whose pair correlations $C(n)$ along $J$-directions
exhibit AF order as opposed to the F-order behavior along the
$J'$-direction. This is respectively shown by the dashed and solid
lines of the inset.}
\end{figure}

\end{document}